# Causality and radiatively induced CPT violation

C. Adam[1], F.R. Klinkhamer[2]

*Institut für Theoretische Physik, Universität Karlsruhe, D–76128 Karlsruhe, Germany*

## Abstract

We consider quantum electrodynamics with an additional Lorentz- and CPT-violating axial-vector term in the fermionic sector and discuss the possibility that radiative corrections induce a Lorentz- and CPT-violating Chern–Simons-like term for the gauge field. From the requirement of causality and the assumed validity of perturbation theory, we conclude that the induced Chern–Simons-like term must be absent in the full quantum field theory.

[1]E-mail address: adam@particle.uni-karlsruhe.de
[2]E-mail address: frans.klinkhamer@physik.uni-karlsruhe.de

Over the last few years, possible consequences of violations of Lorentz and CPT invariance have been actively studied. The Standard Model of known elementary particles and interactions respects both of these symmetries. Signals of Lorentz and CPT violation could, therefore, be indicative of new physics, e.g. quantum gravity [1], superstrings [2], or a new type of CPT anomaly within local quantum field theory [3, 4].

Both phenomenological consequences and theoretical implications of Lorentz and CPT breaking have been studied in a series of papers [5]–[9]. Moreover, there have been extensive discussions on the possibility of Lorentz- and CPT-symmetry breaking in the gauge field sector induced by radiative corrections of an explicitly symmetry-breaking matter sector. See Refs. [7]–[18], where Ref. [18], in particular, contains an extensive list of references. In this Letter, we comment on this last question.

Concretely, we start from the following Lagrangian density:

$$\mathcal{L} = \bar{\psi}\left(i\partial\!\!\!/ - e A\!\!\!/ - m - \gamma_5 b\!\!\!/\right)\psi - \tfrac{1}{4}F_{\mu\nu}F^{\mu\nu} + \mathcal{L}_{\text{gf}}, \tag{1}$$

and use units for which $c = \hbar = 1$. Here, $A_\mu$ is an Abelian gauge potential with field strength tensor $F_{\mu\nu} \equiv \partial_\mu A_\nu - \partial_\nu A_\mu$, $\psi$ a 4-component Dirac spinor with nonzero mass $m$ and $e$ the gauge coupling constant. As usual, the slash of a vector stands for a contraction with the Dirac matrices, $A\!\!\!/ \equiv \gamma^\mu A_\mu$. Further, the quantity $b_\mu$ is a constant, prescribed "four-vector" with the dimension of mass. A gauge fixing term $\mathcal{L}_{\text{gf}}$ has also been added to the Lagrangian (1), since we are interested in the full quantum field theory with dynamical fields $\psi$ and $A_\mu$. The Lagrangian (1) is thus a generalized version of one-flavor quantum electrodynamics (QED) [19], where the generalization is provided by the CPT- and Lorentz-breaking term $-\bar{\psi}\gamma_5 b\!\!\!/\,\psi$.

Let us assume for the moment that we are interested in processes which involve only photons in the initial and final states. Such processes may be described by an effective action which is obtained from Eq. (1) by integrating out the fermionic degrees of freedom,

$$S^{\text{eff}} = \int \mathrm{d}^4 x \left(-\tfrac{1}{4}F_{\mu\nu}F^{\mu\nu} + \mathcal{L}_{\text{gf}} + \ln\frac{\det(i\partial\!\!\!/ - eA\!\!\!/ - m - \gamma_5 b\!\!\!/)}{\det(i\partial\!\!\!/ - m - \gamma_5 b\!\!\!/)}\right). \tag{2}$$

In the limit $b_\mu \to 0$, the effective action (2) is given by the standard effective gauge field action of one-flavor QED [19], whereas for $b_\mu \neq 0$ additional terms will be present.

An example of such an additional term, quadratic in the gauge field, would be a Chern–Simons-like term (see Refs. [5, 20] and references quoted therein),

$$S_{\text{CS-like}} = \int \mathrm{d}^4 x \, \tfrac{1}{4} k_\mu \, \epsilon^{\mu\nu\rho\sigma} A_\nu F_{\rho\sigma}. \tag{3}$$

Here, the quantity $k_\mu$ is a fixed "four-vector" with the dimension of mass. This Chern–Simons-like term (3) has the same behavior under Lorentz and CPT transformations as the term proportional to $b_\mu$ in the Lagrangian (1). It can, therefore, be induced by radiative corrections (concretely, by vacuum polarization). This implies that $S_{\text{CS-like}}$ appears in the effective action (2) with $k_\mu$ related to $b_\mu$,

$$k_\mu = \zeta\, b_\mu, \tag{4}$$

where $\zeta$ is a numerical coefficient that has to be determined. (Terms of higher order in $b_\mu$ turn out to be absent, as will be discussed below.)

The question of how to determine the correct value for this constant $\zeta$ has given rise to a lively debate in the recent literature. The authors of Ref. [7] observed that a calculation which is perturbative in $b_\mu$ leads to a finite but undetermined value for $\zeta$. The authors of Ref. [8], on the other hand, found a vanishing value for $\zeta$ by requiring a gauge-invariant regularization of the axial current, but this requirement has been criticized as being too restrictive from a physical point [10]. In fact, the authors of Ref. [10] again concluded that the value of $\zeta$ remains undetermined. Further calculations in different regularization schemes have since been performed, giving a number of different values for the coefficient $\zeta$. See, for example, Refs. [11]–[18].

In most of the above-quoted articles, the gauge potential is either explicitly treated as an external field or, at least, no use is made of the fact that $A_\mu$ is a dynamical quantized field. The full quantum field theory (1) was studied in Ref. [15], where it was argued that a correct implementation of gauge invariance via Ward identities requires a zero value for $\zeta$. However, this conclusion was again criticized as being too restrictive [18].

The debate reviewed in the last two paragraphs mainly focused on the issue of gauge invariance. It would, therefore, be desirable to have an independent line of reasoning, based on another fundamental property of quantum field theory. The purpose of this Letter is to determine the value of $\zeta$ from the requirement of *microcausality* of the quantum field theory defined by the Lagrangian (1).

Let us, first, recall some results from the literature on the constant $\zeta$ as defined by Eq. (4). One result is that $\zeta$ is a numerical constant that does not depend on the symmetry-breaking four-vector $b_\mu$, provided the term proportional to $b_\mu$ in the theory (1) can be treated perturbatively. Then, the leading contribution to the Chern–Simons-like term (3), which is linear in $b_\mu$, comes from a naively linearly divergent diagram (which is why the contribution is ambiguous, even though it turns out to be finite). All possible higher-order contributions to the Chern–Simons-like term come from convergent diagrams. These contributions are, therefore, unique and arguments based on the classical symmetries of the theory remain valid. From the gauge invariance of the classical axial current one may conclude that these higher contributions to the coefficient $\zeta$ are, in fact, zero. For a more detailed discussion, see, in particular, Refs. [10, 18].

Within perturbation theory, $\zeta$ does not depend on whether $b_\mu$ is spacelike or timelike. This observation will be important in the sequel, because the effective theory (2) with a Chern–Simons-like term (3) included will turn out to behave rather differently for spacelike and timelike $k_\mu$.

From now on, we assume that both $b_\mu$ and $k_\mu$ are purely timelike, that is, $b_\mu = (b_0, 0, 0, 0)$ and $k_\mu = (k_0, 0, 0, 0)$. With this choice, we will find that microcausality can be maintained for the original quantum field theory (1) and, therefore, for the effective field theory (2). It has, however, been demonstrated in Ref. [20] that microcausality is violated for a theory with a Chern–Simons-like term (3) and purely timelike $k_\mu = (k_0, 0, 0, 0)$. This implies that the coefficient $\zeta$ in Eq. (4) should be chosen equal to zero. We like to emphasize at this point that our argument is valid only for a dynamical, quantized gauge field. It does not apply to the external field problem.

Let us then start by investigating the causality behavior of the *original* theory (1) with $b_\mu = (b_0, 0, 0, 0)$. For completeness, we should mention that this theory may have some unusual properties. As pointed out in Section IV B of Ref. [9], a high-energy fermion can, for example, decay by emitting a virtual photon which creates a fermion-antifermion pair ($f \to f + \gamma \to f + \bar{f} + f$). It is, however, not clear that this instability affects the consistency of the theory. For now, we simply assume the validity of the usual perturbation theory methods [19]. The Lagrangian (1) consists of three pieces that have to be investigated separately.

First, the Maxwell and gauge-fixing part of the Lagrangian density (1),

$$\mathcal{L}_{A-\text{kin}} = -\tfrac{1}{4} F_{\mu\nu} F^{\mu\nu} + \mathcal{L}_{\text{gf}}, \tag{5}$$

is known to respect the requirements of causality [19]. Classically, the two polarization modes of the electromagnetic plane wave travel with constant speed $c = 1$. In the quantized Maxwell theory, microcausality holds for physical, gauge-invariant operators (e.g. the transverse components of the gauge potential or the electric and magnetic fields). Concretely, the commutator of two physical operators vanishes for spacelike separations.

Second, the fermionic part of the Lagrangian density (1),

$$\mathcal{L}_{\psi-\text{kin}} = \bar{\psi}\left(i\partial\!\!\!/ - m - \gamma_5 b\!\!\!/\right)\psi, \tag{6}$$

is more difficult. Fortunately, the causal behavior of this quadratic Lagrangian for purely timelike $b_\mu$ has already been determined in Ref. [6]. For the metric $g_{\mu\nu} = \text{diag}(-1,1,1,1)$, the fermionic Lagrangian (6) leads to the dispersion relation

$$(p^2 - b^2 + m^2)^2 + 4p^2 b^2 - 4(p \cdot b)^2 = 0, \tag{7}$$

with the solutions

$$p_0^2 = \omega_\pm^2 \equiv |\vec{p}|^2 + m^2 + b_0^2 \pm 2|b_0||\vec{p}|, \tag{8}$$

for the case of purely timelike $b_\mu = (b_0, 0, 0, 0)$.

The dispersion relation (7) is shown in Fig. 1 for the values $b_0 = m = 1/\sqrt{2}$ (more realistic would, of course, be $|b_0| << m$). Two remarks are in order. First, the dispersion relation (7) has four real roots $p_0 = q_i$, $i = 1\ldots 4$, for arbitrary values of $|\vec{p}|$. Second, the group velocities,

$$\vec{v}_g^\pm \equiv \frac{\partial \omega_\pm}{\partial \vec{p}} = \frac{\vec{p}}{|\vec{p}|} \frac{|\vec{p}| \pm |b_0|}{\sqrt{(|\vec{p}| \pm |b_0|)^2 + m^2}} \leq 1, \tag{9}$$

do not exceed the constant velocity of light $c$, which is equal to 1 in our units. This classical reasoning indicates that the causal structure of the theory (6) remains unchanged by the inclusion of a purely timelike $b_\mu$.

In the corresponding quantum field theory, the issue of microcausality is determined by the anti-commutator function $iS(x-y) \equiv \{\psi(x), \bar{\psi}(y)\}$, which has to vanish for spacelike separations $(x-y)^2 > 0$ in order to maintain microcausality. This anti-commutator has been calculated in Ref. [6] for a purely timelike $b_\mu = (b_0, 0, 0, 0)$. The result is that $S(x-y)$ may be expressed like

$$S(x-y) = (i\partial\!\!\!/ - \gamma_5 b\!\!\!/ + m)(i\partial\!\!\!/ + \gamma_5 b\!\!\!/ + m)(i\partial\!\!\!/ + \gamma_5 b\!\!\!/ - m)\,\Delta(x-y), \tag{10}$$

where $\Delta(x)$ is defined as

$$\Delta(x) = i \oint_C \frac{dp_0}{2\pi} \int \frac{d^3 p}{(2\pi)^3} \frac{e^{ip\cdot x}}{(p^2 - b^2 + m^2)^2 + 4p^2 b^2 - 4(p\cdot b)^2}. \tag{11}$$

Here, the integration contour $C$ encircles all four poles in the counter-clockwise direction. For $b_\mu = (b_0, 0, 0, 0)$, the integration can be performed explicitly [6]:

$$\Delta(x) = -\frac{i}{8\pi} \frac{x^0}{|x^0|} \frac{\sin b_0 |\vec{x}|}{b_0 |\vec{x}|} J_0(m\sqrt{-x^2}), \tag{12}$$

for timelike separation $x^2 < 0$ ($J_0$ is the zero-order Bessel function), and

$$\Delta(x) = 0, \tag{13}$$

for spacelike separation $x^2 > 0$. Hence, microcausality is maintained for the theory (6) with purely timelike $b_\mu$.

The last part of the original Lagrangian density (1) that has to be studied is the interaction term,

$$\mathcal{L}_{A\psi-\text{int}} = -e\bar{\psi}\slashed{A}\psi. \tag{14}$$

This term is strictly local, so that causality holds at the classical level. However, the composite operator $J_\mu = \bar{\psi}\gamma_\mu\psi$ is singular in the quantum theory and needs regularization. Treating the interaction term perturbatively, some calculations may then lead to ambiguous results. These ambiguities are fixed by the symmetries of the theory and by the appropriate physical conditions. One physical condition is certainly that causality should continue to hold for the full quantum field theory, if at all possible. Therefore, microcausality is a requirement at the level of the perturbatively defined quantum field theory (1).

Next, we turn to the causality behavior of the *effective* theory (2), with the Chern–Simons-like term (3) included. This problem has already been studied in Ref. [5] for the classical field theory and in Ref. [20] for the quantum theory. Here, we only need to review the results. The quadratic part of the effective action reads

$$S_{A-\text{kin}}^{\text{eff}} = \int d^4x \left( \tfrac{1}{2} A_\mu \left( g^{\mu\nu}\Box - \partial^\mu\partial^\nu - \epsilon^{\mu\nu\rho\sigma}k_\rho\partial_\sigma \right) A_\nu + \mathcal{L}_{\text{gf}} \right), \tag{15}$$

with $k_\mu = (k_0, 0, 0, 0)$. Ignoring the gauge-fixing part for the moment, one finds the dispersion relation [5]

$$p^4 + k^2 p^2 - (k \cdot p)^2 = 0, \tag{16}$$

with the solutions

$$p_0^2 = \omega_\pm^2 \equiv |\vec{p}|^2 \pm |k_0||\vec{p}|. \tag{17}$$

The dispersion relation (16) is shown in Fig. 2. It is obvious that there is no separation into positive and negative frequency parts. In addition, the group velocities of both degrees of freedom may become arbitrarily large. For the plus sign in Eq. (17), one has, for example,

$$\vec{v}_g^{\,+} \equiv \frac{\partial \omega_+}{\partial \vec{p}} = \frac{\vec{p}}{|\vec{p}|} \frac{2|\vec{p}| + |k_0|}{2\sqrt{|\vec{p}|^2 + |k_0||\vec{p}|}} \,, \tag{18}$$

which becomes arbitrarily large for arbitrarily small $|\vec{p}|$, as long as $|k_0| \neq 0$. For the minus sign in Eq. (17), the energy even becomes imaginary at low momenta $|\vec{p}| < |k_0|$. In the quantized theory, these low-momentum modes would grow exponentially with time (instead of evolving unitarily). It is, therefore, mandatory to exclude these modes from the theory in order to maintain unitarity. This, in turn, leads to a violation of microcausality in the theory (15), as has been shown in Section 5.2 of Ref. [20]. For a purely spacelike $k_\mu$, on the other hand, there is no violation of microcausality, see Section 5.1 of Ref. [20].

To summarize, we have studied the issue of microcausality for the quantum field theory (1). We have found that both the purely bosonic part (Maxwell theory) and the purely fermionic part (Dirac theory with an additional $b_\mu$-term) maintain microcausality. Therefore, the free theory (1) without interaction term (14) is causal. If we now impose the requirement that microcausality continues to hold for the full quantum field theory (1) with the interaction term (14) included, we conclude that the Chern–Simons-like term (3) must be absent in the effective action (2) for purely timelike $b_\mu$. The reason is that microcausality would be violated by the Chern–Simons-like term for purely timelike $k_\mu$, at least within the realm of unitary perturbation theory. For general timelike $b_\mu$, there always exists a coordinate system for which $b_\mu$ is purely timelike. Therefore, our argument holds also for general timelike $b_\mu$.

For spacelike $b_\mu$, we assume moreover that a perturbative expansion in $b_\mu$ is possible for sufficiently small values of $b_\mu$, that is, the coupling constants in the effective gauge field action (2) can be expanded in powers of $b_\mu$. (The effective expansion parameter is $b^2/m^2$, on dimensional grounds.) With this additional assumption, the Chern–Simons-like term (3) must also be absent in the effective action (2) for sufficiently small spacelike $b_\mu$.

In other words, the ambiguity of the value of the coefficient $\zeta$, which is present in a naive perturbative treatment, is fixed by the requirement of microcausality of the full quantum field theory (1). The unique value for the coefficient $\zeta$ of Eq. (4) which preserves microcausality is

$$\zeta = 0 \,. \tag{19}$$

We emphasize once more that the calculations were performed for the specific choice of a purely timelike $b_\mu$. But, within perturbation theory, the value of the coefficient $\zeta$ does not depend on $b_\mu$, so that $\zeta = 0$ holds for arbitrary $b_\mu$. Note also that the well-known Pauli–Villars and dimensional regularization methods do indeed give $\zeta = 0$, at least for small enough $b^2/m^2$; cf. Refs. [7, 15].

As a consequence, the Lorentz and CPT violation introduced by the $b_\mu$ term in the Lagrangian (1) does not appear in the effective action (2) via the quadratic local Chern–Simons-like term (3). It is, of course, still possible that Lorentz and CPT violation is induced in the effective action (2) by higher-order local terms or even nonlocal terms.

At this moment, it may be of interest to mention a recent calculation of the induced Chern–Simons-like term (3) at finite temperature [21], which agrees with our general argument. Irrespective of the ambiguity of the coefficient $\zeta$ at zero temperature, the authors of Ref. [21] find a unique and finite additional contribution to the induced Chern–Simons-like term at finite temperature. This finite-temperature contribution is nonzero for the spacelike component of $b_\mu$ : $k_i^{\text{finite}-T} = 16\, F(\xi)\, b_i$, with $\xi \equiv m/2\pi T$ and the function $F(\xi)$ as calculated in Ref. [21]. (Note that $F$ vanishes as the temperature $T$ goes to zero.) On the other hand, no extra Chern–Simons-like term is induced for the time component $b_0$ : $k_0^{\text{finite}-T} = 0$. The heat bath is, of course, a further source of explicit violation of Lorentz invariance, which explains the different results for timelike and spacelike directions. But these finite-temperature results are precisely what one expects from the general requirement of causality. A Chern–Simons-like term with a purely spacelike $k_i$ does not lead to a violation of microcausality [20] and its presence is compatible with the requirement of causality. On the other hand, a nonzero induced $k_0$ would violate causality and must therefore be absent.

One physical consequence of a nonvanishing spacelike $k_\mu$ is the birefringence of light *in vacuo*. Observations of distant radio galaxies severely restrict the possible values of the $k_\mu$ parameter: $|k_\mu| \lesssim 10^{-33}$eV, see Refs. [5, 22] and references quoted therein. If $\zeta$ had a unique nonzero value of order $\alpha \equiv e^2/4\pi$, this bound would translate into a strong upper bound on the Lorentz- and CPT violating parameter $b_\mu$ for the fermion sector, as pointed out in Refs. [7, 8, 10]. Our result $\zeta = 0$ implies that no such strong experimental upper bound on $b_\mu$ exists, leaving room for a comparatively large Lorentz and CPT violation in the fermionic matter sector (provided the $b_\mu$-like terms do not affect the consistency of the theory).

If it should turn out that, after all, $k_\mu$ has a tiny nonzero value, then the result $\zeta = 0$ implies that this nonzero value cannot be traced back to a symmetry-breaking term in the matter sector for Dirac fermions, at least for a Lagrangian like Eq. (1). A nonzero Chern–Simons-like term (3) could then be the low-energy remnant of a more fundamental theory like superstring theory, via a mechanism that is not yet understood in detail [2]. Or, it could be induced by Weyl fermions over a nontrivial spacetime manifold via the CPT anomaly [3, 4].

In fact, it has been shown in Ref. [3] that certain chiral gauge theories, defined over a four-dimensional spacetime manifold with a compact separable dimension and appropriate spin structure, give rise to a CPT anomaly precisely of the form of the Chern–Simons-like term (3). (An example of an appropriate chiral gauge theory would be the well-known $SO(10)$ grand-unified theory with three families of quarks and leptons or even the Standard Model with an additional assumption, see Section 5 of Ref. [3].) In this case, the magnitude of $k_\mu$ is given by the inverse of the length of the compact dimension, thereby naturally accounting for the required smallness of $|k_\mu|$. The direction of $k_\mu$ is given by the direction of the compact dimension, so that the different causality behavior of timelike and spacelike $k_\mu$ does not come unexpected [20]. Indeed, a spacetime manifold with a compact timelike dimension contains closed timelike curves and has causality problems already at the macroscopic level.

# Acknowledgements

We thank R. Jackiw and M. Perez-Victoria for useful comments.

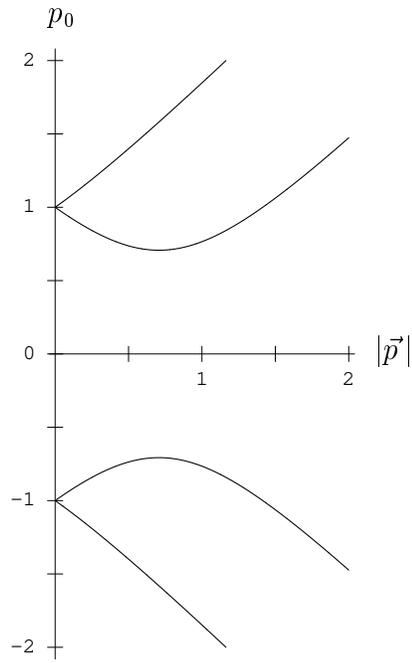

Figure 1: The dispersion relation (7) in the $(p_0, |\vec{p}|)$ halfplane for purely timelike parameter $b_\mu = (m, 0, 0, 0)$ and fermion mass $m = 1/\sqrt{2}$.

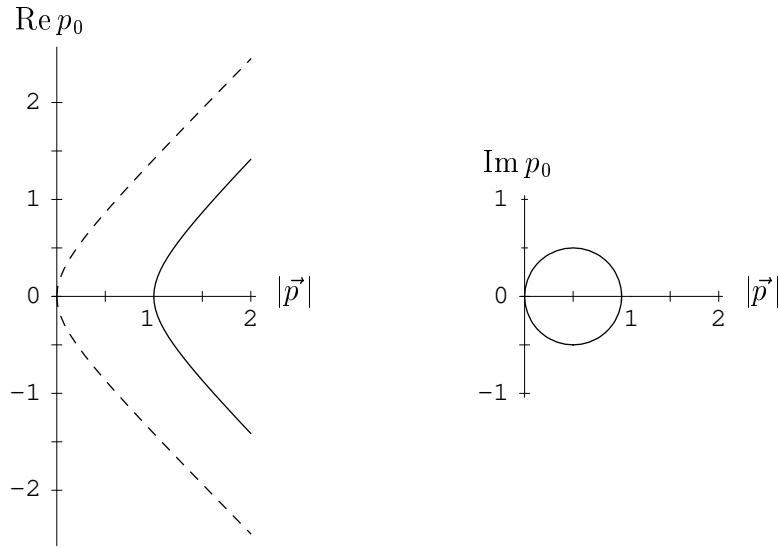

Figure 2: The dispersion relation (16) in the $(\operatorname{Re} p_0, |\vec{p}|)$ and $(\operatorname{Im} p_0, |\vec{p}|)$ halfplanes for purely timelike Chern–Simons parameter $k_\mu = (1, 0, 0, 0)$, with broken (solid) curves corresponding to the plus (minus) sign in Eq. (17).

11